\documentclass[onecollarge]{svjour2}       
\smartqed  
\usepackage{graphicx}
\usepackage{color}
\begin{document}

\title{1/f noise in  turbulent flows} 
 \subtitle{ transitions with heavily tailed distributed interevent durations.}


\author{J. Herault \and  F. P\'etr\'elis \and S. Fauve  }


\institute{Helmholtz-Zentrum Dresden-Rossendorf, Bautzner Landstr. 400, 01328 Dresden, Germany
           \and
        Laboratoire de Physique Statistique, Ecole Normale Sup\'erieure, CNRS, Universit\'e P. et M. Curie, Universit\'e Paris Diderot, Paris, France
}

\date{Received: date / Accepted: date}

\maketitle

\begin{abstract}
 {We report the experimental observation of} $1/f$ fluctuations  {in three different turbulent flow configurations:} the large scale velocity driven by a two-dimensional turbulent flow, the magnetic field  {generated by} a turbulent swirling flow of liquid sodium and the pressure fluctuations due to vorticity filaments in a swirling flow. For these three systems, $1/f$ noise is shown to result from the dynamics of coherent structures that display transitions between a small number of states. The interevent duration is distributed as a power-law. The  exponent of this power-law and the nature of the dynamics (transition between symmetric states or asymmetric ones) select the exponent of the $1/f$ fluctuations. 

\end{abstract}
\section{Introduction}
\label{intro}

In the past fifty years, a lot of progress has been made in the characterization of coherent structures and their impact on the statistical properties of turbulent velocity fields \cite{Hussain86}. Some studies have reported the presence of long range correlations related to $1/f$ noise in the frequency power spectrum of turbulent fluctuations.  $1/f$ noise has been observed in the solar wind \cite{Matthaeus86}, von K\'arm\'an flows \cite{Abry,Ravelet}, magnetohydrodynamic  flows \cite{Bourgoin,Ponty,Monchaux,Dmitruk} and two-dimensional turbulence \cite{Dmitruk,Herault}.  It has been shown in these studies that  $1/f$ noise is related to the slow dynamics of coherent structures but the explicit relation between the spectrum and the dynamics has been emphasized only recently \cite{Herault}.
Originally observed in solid state physics \cite{VanderZiel},    $1/f$ noise  (also known as flicker noise) refers to a self-similar power spectrum with an exponent close to $-1$. This definition has been extended to power  spectra  of the form $E(f) \propto f^{-\alpha}$, with  $\alpha$ between $0$ and $2$. Such exponents are not unusual in turbulence but the frequency range over which they are observed is. 
The standard description of turbulent cascade is that velocity fluctuations have energy   between the scale of energy injection $l_I$ and  the scale of energy dissipation $l_d$.  The corresponding temporal range is limited  by $\tau_I$ and $\tau_d$, the turn-over time of eddies of size $l_I$ and $l_d$. However,  $1/f$ spectrum in turbulence is observed for frequencies  much smaller than $\tau_d^{-1}$ and $\tau_I^{-1}$. In other words, no straightforward relation exists between the frequencies  and the spatial wave numbers and the usual tools, like the Taylor's hypothesis or dimensional analysis, fail to predict the value of the exponent $\alpha$. Indeed, we will show that the origin of  $1/f$ noise  is  the  dynamics of  coherent structures with lifetimes much larger than their turn-over  time.

In the context of deterministic dynamical systems, $1/f$ fluctuations were initially discussed by Manneville \cite{Manneville80}  {using} a  {mapping} which exhibits long laminar phases interrupted by chaotic bursts.  {$1/f$ noise} is  {related} to a self-similar distribution of waiting times between bursts. More precisely, a signal made of bursts  {with waiting times} $\tau$ distributed as $\tau^{-2}$, exhibits a $f^{-1}$ spectrum. In a similar approach, Geisel et al \cite{Geisel87} showed  {that} the power spectrum is given by $E(f) \sim f^{-\alpha}$ for a distribution of form $P(\tau) \sim \tau^{\alpha-3}$. Lowen and Teich \cite{Lowen} extended  {this} result  {to the case of random transitions} between symmetric states.  

It is tempting to apply the previous approach to coherent structures coexisting with turbulent fluctuations. This requires to identify  transitions between  {different} states.  {It has been shown that} two-dimensional turbulence  ({respectively von} K\'arm\'an flows) display  transitions between symmetric states \cite{Sommeria86,Gallet}  ({respectively} between different flow patterns \cite{Ravelet}).  {We show here} that the statistical properties of these transitions are responsible for the $1/f$ fluctuations. We consider three different quantities: the large scale circulation in two-dimensional turbulence, the magnetic field  {generated by a} von K\'arm\'an  {flow of liquid sodium} and the  {wall} pressure  {fluctuations} in a turbulent von K\'arm\'an  {swirling} flow. For all these systems, we demonstrate that $1/f$ noise is  {related to} the dynamics of large scale coherent structures that transition between different states. The  distribution of the time spent in one of the states follows a power-law with an exponent that sets the value of the exponent of the power spectrum. 

{The organization of the paper is as follows. In section II the renewal theory and the forecast of Lowen and Teich \cite{Lowen} are  {recalled}. A qualitative argument is presented to understand the connection between the $1/f$ spectrum and self-similar distribution  {of waiting times}. In section III, we  {analyze the experimental data obtained in the previously mentioned configurations in the light of these theoretical predictions}. 
}


\section{Renewal process with heavily tailed distributions of interevent durations}
%
\subsection{Description}

\begin{figure}[htb!]
\begin{center}
\includegraphics[width=110mm,height=60mm]{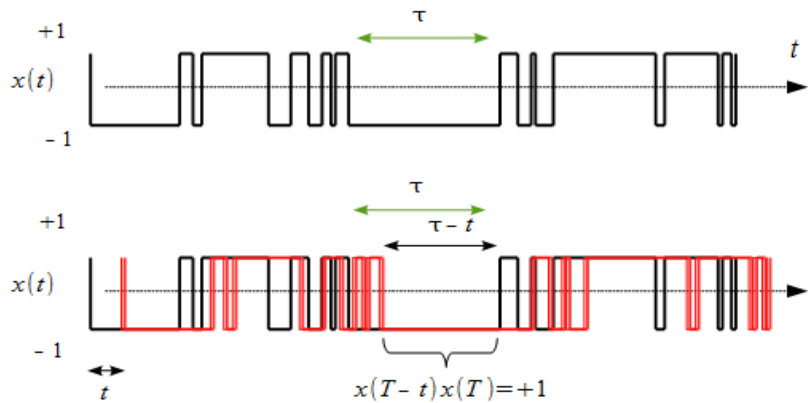}
\caption{Top: renewal process defined  by random transitions between two-states, associated to the values $x=\pm 1$. Bottom : the auto-correlation function $C(t)$ is given by the integration of the product $x(T)x(T-t)$, which is mostly made of long phases of constant polarity and phases of fast oscillations. The former phases contribute to the long-range correlation.}
\label{fig1}
\end{center}
\end{figure}

A renewal process is a stochastic process defined by a sequence of $N$ events  associated to a  {series} of durations $(\tau_i)_N$, with $\tau$  random, positive and independent identically distributed variables. In the following, the events correspond to transitions between two states $S$ called $A$ and $B$ and $\tau_i$ is the time spent in one state after the $i-1$  {event}. We introduce the variable $x(t)$, such that $x(t)=x_A$, when $S=A$ and $x(t)=x_B$, when $S=B$. The distribution of the duration $\tau$  is $P_A$ for the state $A$  {respectively $P_B$ for $B$}. A process is  symmetric when $P_A=P_B$. Figure \ref{fig1} (top) illustrates a  symmetric renewal process with $x_A=1$ and $x_B=-1$. 

The power spectrum of $x$ is defined by

\begin{equation}
E(f)=\frac{1}{T_f} \left\langle \hat x(f)  \hat x^{*}(f) \right\rangle \quad \hbox{with} \quad \hat x(f)= \int^{T_f}  _{0} x(t) e^{i 2 \pi f t} dt 
\end{equation}

\noindent where  $\left\langle \cdot \right\rangle$ is the average over the realizations and $T_f=\sum^N _{i=1} \tau_i$ is the duration of the process,  {that ultimately tends to infinity}. In this study, we are interested in self-similar power spectrum with $E(f) \sim f^{-\alpha}$, and $0<\alpha<2$.


\subsection{Relation between the exponents $\alpha$ and $\beta$}

We start with  a qualitative  {argument to explain} how the distribution $P(\tau) \sim \tau^{-\beta}$ controls the value of the exponent $\alpha$ \cite{Lowen}.  To wit, we first obtain the auto-correlation $C(t)$ of $x$ and then we calculate the power-spectrum of $x$ using the Wiener-Khinchin theorem. The autocorrelation function $C(t)$ is defined by

\begin{equation}
C(t)= \left\langle   x(0) x( t)   \right\rangle
\end{equation}
 
The Wiener-Khinchin theorem states that the Fourier transform of $C(t)$ converges to $E(f)$ for $T_f \rightarrow \infty$.  We consider a symmetric process with  $x_A=1$ and $x_B=-1$ as sketched in  figure \ref{fig1} (bottom). We fix $\beta>2$,   in order to consider only   processes with   $ \langle \tau \rangle < \infty$. For an ergodic process and $T_f \gg \langle \tau \rangle$, the auto-correlation $C(t)$  is obtained by
\begin{equation} 
C(t)   =    \frac{1}{T_f}\int_0 ^{T_f}   x(T-t) x(T) dT 
\end{equation}

We observe that the product $x(T) x(T-t)$ is composed of fast oscillations and  long periods of constant polarity, due to long phases of duration $\tau$ in $x(T)$. We assume that only the phases with $\tau>t$ contribute to the autocorrelation with a contribution $\tau-t$. In other words, the average contributions of short phases with $\tau<t$ vanish.  It follows that the autocorrelation is well approximated by

\begin{equation}
C(t){ \simeq \frac{1}{T_f}\int_t ^{T_f}  ( \tau-t) n( \tau) d \tau \quad \hbox{for} \quad  \langle \tau \rangle \ll t \ll T_f } 
\label{eqC1}
\end{equation}
with $n(\tau)$ the number of phases of duration $\tau$, which  is equal to $P(\tau) T_f/ \langle \tau \rangle$, with $T_f/ \langle \tau \rangle$ the total number of events.  Then,  equation  (\ref{eqC1}) becomes

\begin{equation}
C(t) \simeq \frac{1}{\langle \tau \rangle}\int_t ^{T_f}  ( \tau-t) P( \tau) d \tau .
\end{equation}

\noindent If $P(\tau)$ is an exponential function, the autocorrelation function is also an exponential  function, as expected for a Poisson process. For $P(\tau) \sim \tau^{-\beta}$ and $\beta>2$,  the autocorrelation scales as $C(t)\sim t^{-\beta+2}$. Finally, the power spectrum $E(f)$ is given by the Fourier transform of $C(t)$

\begin{equation}
E(f) \sim f^{\beta-3} \int u^{-\beta+2} e^{-2 \pi u i} d u \quad 
\end{equation}

\noindent with the change of variable $u=ft$ and for $T_f \rightarrow \infty$. We thus obtain $\alpha=3-\beta$. The result is extended to $1 < \beta < 2$  \cite{Lowen,Niemann} by considering a distribution with $P \sim \tau^{-\beta}$ for $\tau_i \ll \tau \ll \tau_e$, and zero otherwise or exponentially distributed. For a symmetric process ($P_A=P_B$), the power spectrum is then given by

\begin{equation}
E(f) \sim \left\{
\begin{array}{ll}
f^{-(3-\beta)}&   \hbox{for} \quad 1<\beta<3 \\ 
\ln(\tau_i f) & \hbox{for} \quad \beta=3 \\  
\end{array} \right.
\label{symAlpha}
\end{equation}

\noindent for $\tau_e^{-1} \ll f \ll \tau_i^{-1}$ .

 \begin{figure}[htb!]
\begin{center}
\includegraphics[width=110mm,height=30mm]{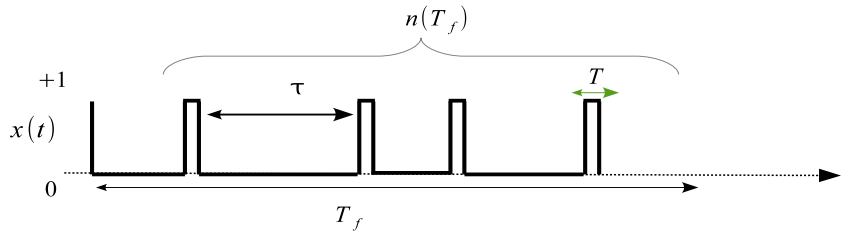}
\caption{Time series displaying bursts from $x=0$ to $x=1$. The process is asymmetric and the duration of the low amplitude phases is distributed according to a power-law, $P\sim \tau^{-\beta}$. In contrast the large amplitude phases have durations that are not distributed following a heavy tailed distribution.}
\label{fig1new}
\end{center}
\end{figure}

Bursting processes have been    considered. The signal $x(t)$  is composed of short intermittent bursts separated by intervals $\tau$ distributed as $P\sim \tau^{-\beta}$ (see figure \ref{fig1new}). The result differs from the symmetric case when $\beta < 2$. Consider a time series for which $x(0)=1$. Let $n(T_f)$ be the number of bursts  {within a time interval of length} $T_f$. We estimate 
\begin{equation}
S=\int_0^{T_f}x(0) x(t) dt\simeq n(T_f) T
\end{equation}  
where $T$ is the duration of a bursts. The number of burst is roughly, $n(T_f)=T_f/\langle \tau \rangle_{T_f}$ where $\langle  \tau \rangle_{T_f}$ is the average duration of the $x=0$ phases when we consider a time series of duration $T_f$. For $\beta < 2$,  $\langle \tau \rangle_{T_f}$ does not tend to a constant at large $T_f$, which obviously results from the divergence of the first moment of the distribution $P$. Then an estimate of $\langle \tau \rangle_{T_f}$ is given by 
\begin{equation}
\langle \tau \rangle_{T_f}=\int_0^{T_f} \tau P(\tau) d\tau \simeq T_f^{2-\beta}
\end{equation} 
We thus have $n(T_f)\simeq T_f^{\beta-1}$. Averaging $S$ over realizations and  {differentiating} with respect to $T_f$, we obtain the autocorrelation function as $C(t) \propto t^{\beta-2}$. The exponent of the power spectrum then satisfies $\alpha=\beta-1$.

To sum up, the predictions for  the power spectrum of a bursting process are \cite{Lowen,Niemann} 

\begin{equation}
E(f) \sim \left\{
\begin{array}{ll}
f^{-(\beta-1)}&   \hbox{for} \quad 1<\beta<2 \\ 
\ln(f\tau_i)^{-2} f^{-1}&   \hbox{for} \quad \beta=2 \\
f^{-(3-\beta)}&   \hbox{for} \quad 2<\beta<3 \\
\ln(\tau_if) & \hbox{for} \quad \beta=3 
\end{array} \right.
\label{asymAlpha}
\end{equation} 

We note that for symmetric or bursting processes, $\alpha$ is equal to $3-\beta$ for $\beta>2$. It implies that the effect of asymmetry in the process  {is} relevant  only for distributions with $\beta<2$. 

These results hold for a  range of frequencies $f$ corresponding to the inverse range of durations $\tau$. In particular, if $P(\tau)$ displays a cut-off for time larger than $\tau_e$,  $1/f$ noise may be observed down to the low cutoff frequency $\tau_e^{-1}$. 
This property allows  {us} to consider $1/f$ noise with low frequency cut-off  $\tau_e^{-1}$, related to an exponential decay of $P(\tau)$ for $\tau \gg \tau_e$. 

\section{Application to turbulent time series}

We now illustrate how  coherent structures  generate   {a} $1/f$ spectrum in three different  {configurations} involving turbulent flows. After  {shortly describing} the experiments, we identify transitions in the  {time series}.  {We then} show that the measured exponents $\alpha$ of the power-spectrum and $\beta$ of the distribution of waiting  {times} follow the theoretical predictions of  section $2$, which depend on the value of $\beta$ and on the nature of the process (bursting or symmetric).  We recapitulate the results in the following table.
 
\begin{table}[htb!]
\label{table:TabE}

 \begin{center}
\begin{tabular}{|c|c|c|c|c|c|c|}
\hline
\rule[-1ex]{0pt}{4ex} \textbf{System} & \textbf{Variable} & \textbf{Process}  & $\alpha$ & $\beta$ & \textbf{Relation} \\
\hline
\rule[-1ex]{0pt}{4ex} Two-dimensional turbulence & flow rate & symmetric &  $0.7$ & $2.25$ &$ \alpha=3-\beta$ \\
 \hline
\rule[-1ex]{0pt}{4ex} Von K\'arm\'an Sodium Dynamo& magnetic field& bursting & ~$0.5$ & $2.5$ &$ \alpha=3-\beta$ \\
 \hline
\rule[-1ex]{0pt}{4ex} Von K\'arm\'an flow & pressure & bursting & $~0.6$ & ~$1.58$  &$ \alpha=\beta-1$ \\
\hline 
\end{tabular}

\vspace{0.15cm}

\caption{Summary of the experimental results.  {We recall the experimental configuration (first column), the measured variable (second column), the nature of the process (third column), the measured exponents $\alpha$ and $\beta$ and the expected relation between them, depending on the value of $\beta$ and on the nature of the process.}  }
\end{center}
\end{table}


\subsection{Two-dimensional turbulence}
 
\begin{figure}[htb!]
\begin{center}
\hbox{
\includegraphics[width=75mm,height=50mm]{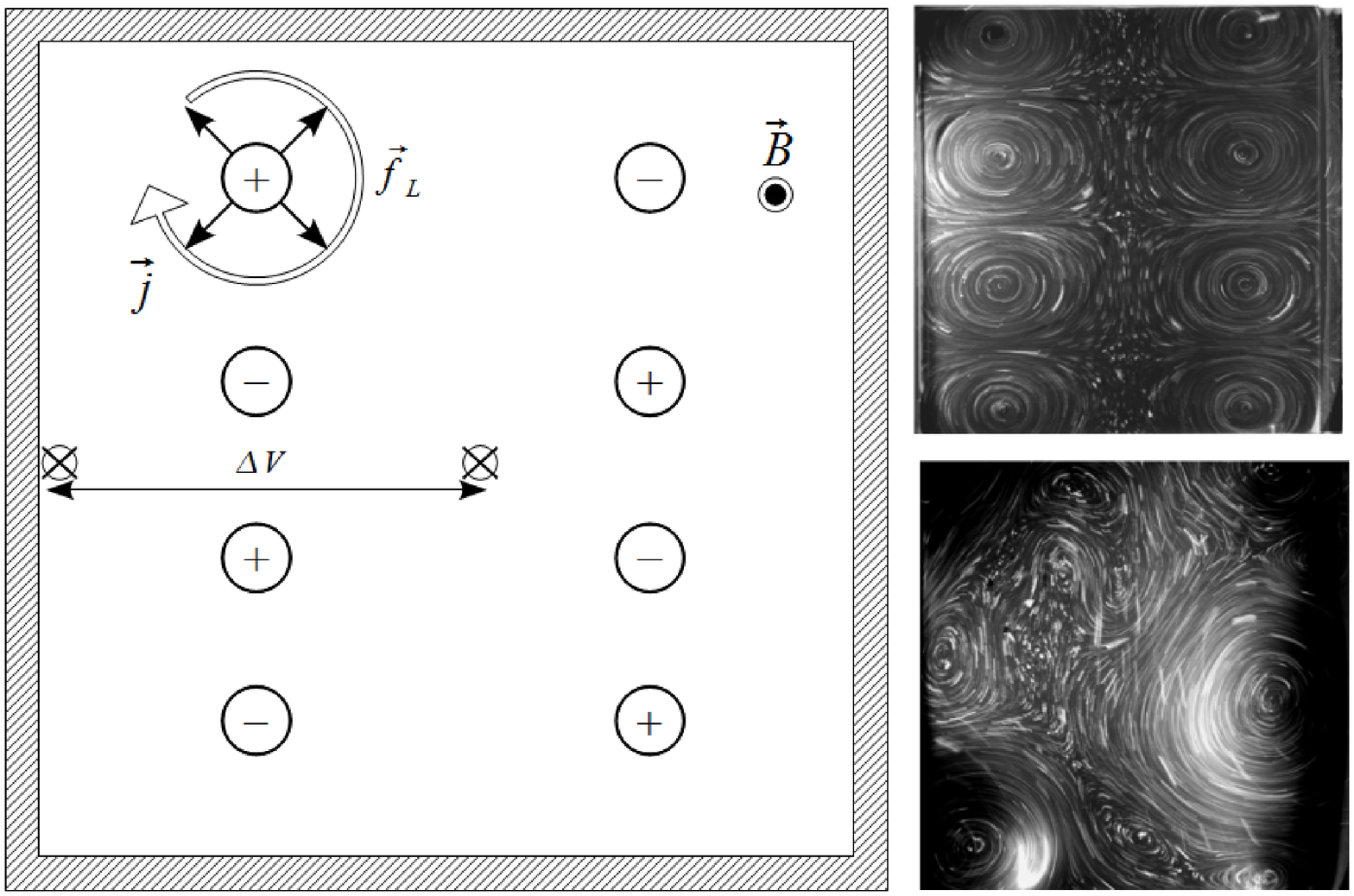} 
\includegraphics[width=70mm,height=50mm]{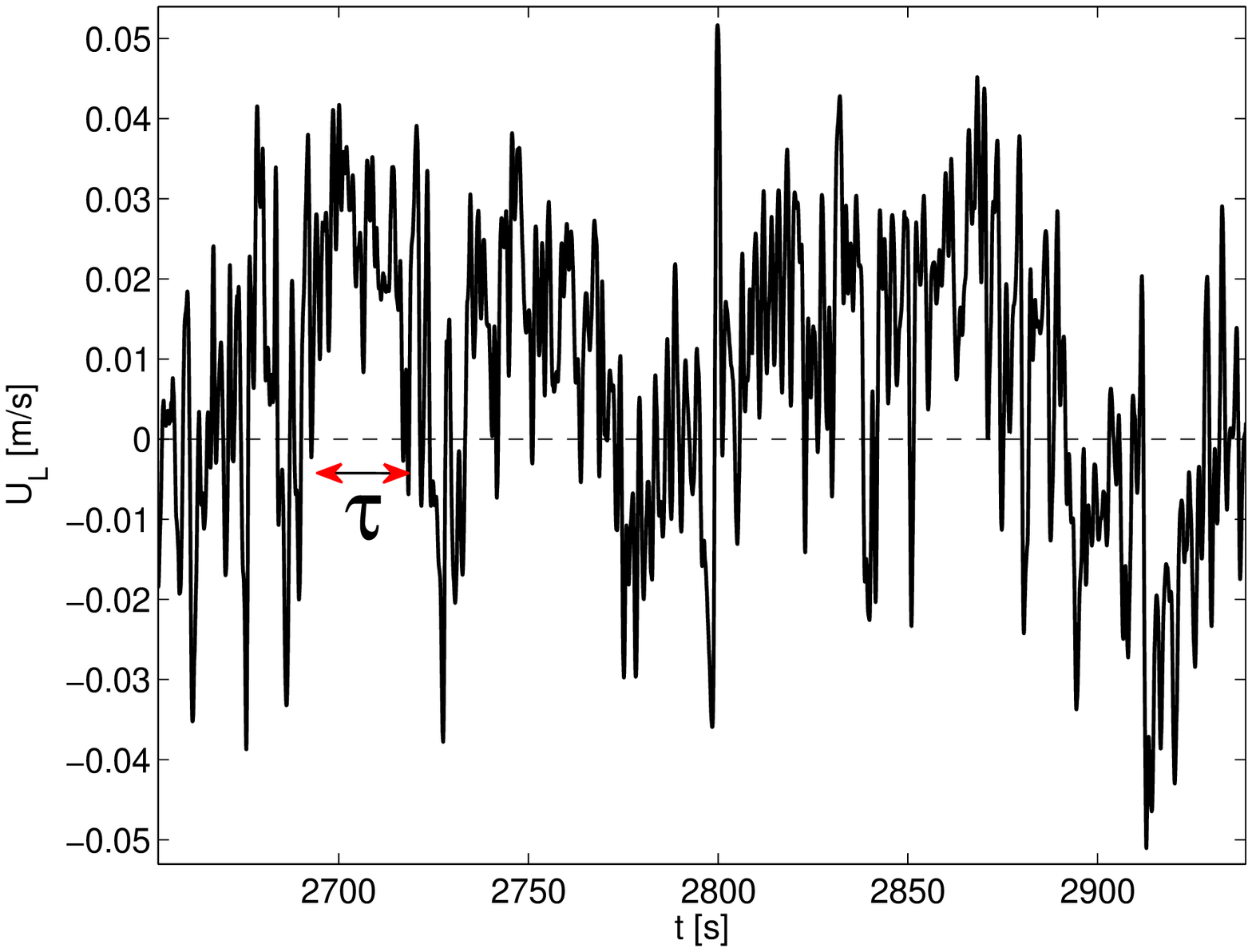}}
\end{center}
\caption{Left: Schematics of the experimental set-up. The Lorentz-force close to the electrodes generates an array of height counter-rotating vortices in the laminar regime (center top picture). The turbulent flow (center bottom picture) is characterized by coherent vortices forcing a large scale circulation. The flow rate  is measured between the center and the wall, using the induced voltage $\Delta V$. Right: time series $ U_L(t)$ of the flow rate divided by the distance $L/2$ for $Rh=15$.}
\label{fig_2D_turb}
\end{figure}

The experiment consists in driving a flow with an electromagnetic force in a conducting fluid, as  described in Refs. \cite{Sommeria86}. A thin layer of  liquid metal (Galinstan) of thickness $h=2cm$, is contained in a square cell of length L= 12 cm.  The cell is located   {inside} a coil producing a uniform vertical  {magnetic} field of strength $B_0 =0.98 \hbox{T}$. A DC current $I$ (0-200A) is  {driven} through the bottom of the cell using a periodic array of 8 electrodes  {with alternate polarities}. The density current $\textbf j$ is radial close to each electrode. The  Lorentz force $\textbf f_L=\textbf j \times \textbf B_0/\rho$, with $\rho$ the density, creates locally a torque. For small injected currents, the laminar flow corresponds to an array of 8  {counter-rotating} vortices shown in figure \ref{fig_2D_turb} (top,  {center}). The flow remains mostly two-dimensional, due to  {Ohmic dissipation} of the perturbations  {with velocity dependence along the direction of the magnetic field.}

When the forcing is strong enough, the flow  becomes turbulent (Fig.\ref{fig_2D_turb}, left). Two-dimensional turbulence is characterized by an inverse cascade of energy. The energy is transferred  to  large scales and coherent structures display lifetimes larger than their  turn-over time.    The dissipation is mostly provided by the friction of the bulk flow with the bottom boundary layer for very large Reynolds number $Re$. The ratio between the forcing and the dissipation  {that results from bottom friction} is given by the dimensionless number $Rh$. Typically, the flow becomes turbulent for $Rh \simeq 5$ and coherent structures appear for $Rh \simeq 12$.  

The velocity measurements are performed with a pair of electric potential probes (Fig. \ref{fig_2D_turb}): one is located in the middle of the cell and the other one close to the lateral wall. The probes measure a potential difference $\Delta V$, due to the integral contribution of the local electromotive force $E=- \textbf u \times \textbf B_0$. Thus, $\Delta V= \phi_L B_0$, with $\phi_L$ the flow rate between   the center and the wall of the cell. We  use the  velocity amplitude $U_L$, defined by $U_L=\phi_L/(L/2)$. Due to flow rate conservation, the velocity amplitude $U_L$ is also equal to the averaged azimuthal velocity. The   voltage is sampled at a frequency rate $f_r=10\hbox{Hz}$. A typical time series $U_L(t)$ is shown  in fig. \ref{fig_2D_turb} (right).

\begin{figure}[htb!]
\begin{center}
\hbox{
\includegraphics[width=70mm,height=50mm]{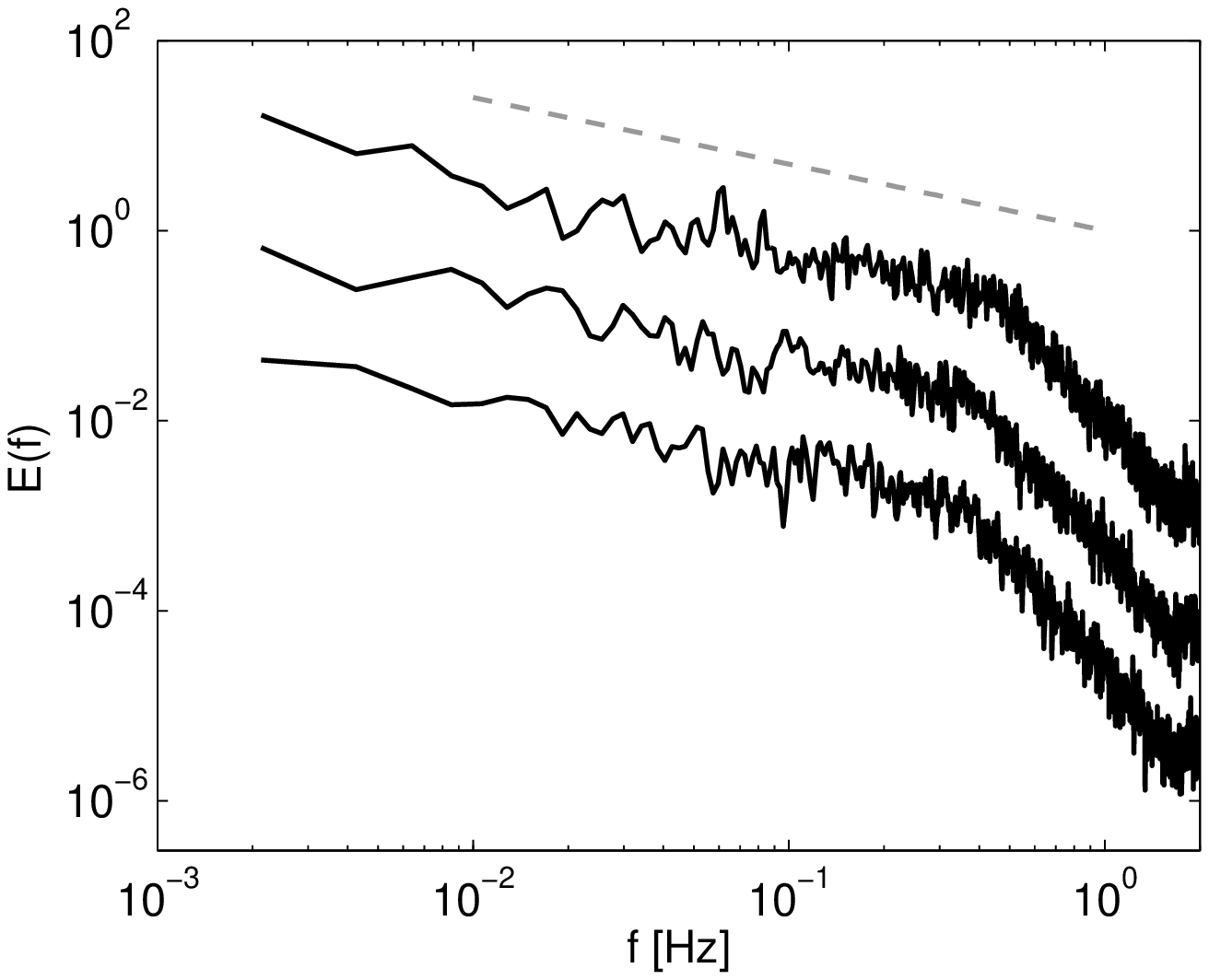}
\includegraphics[width=70mm,height=50mm]{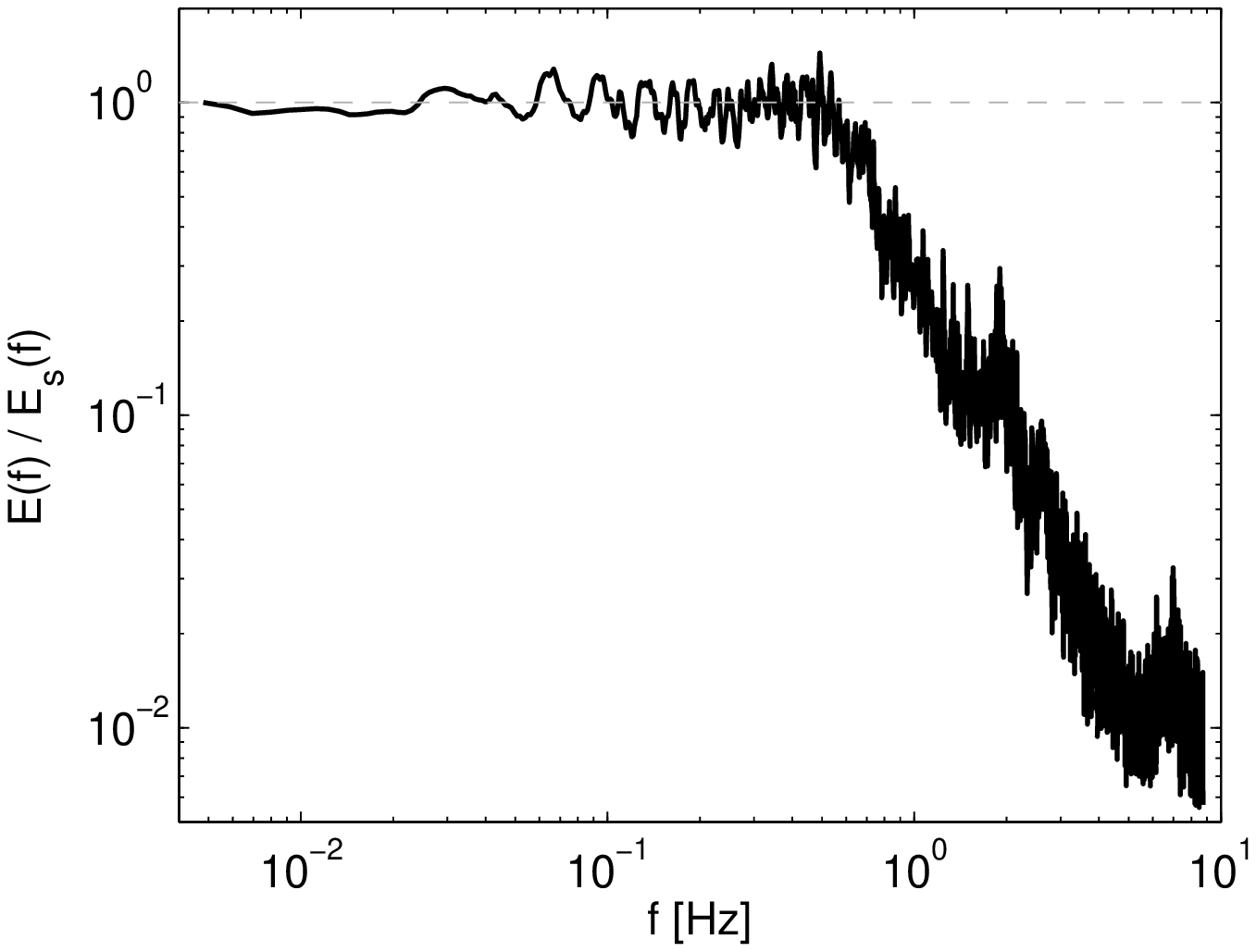} 
}\end{center}
\caption{Left: frequency power spectra $E(f)$ of $U_L(t)$ for  $Rh=  16 ,   19  ,  24$ (from bottom to top). For clarity, the  {spectra  have been multiplied by $1$, $10$ and $100$}. The $f^{-0.7}$ law is displayed as a dashed line. Right: ratio of the  power spectra of   $s(t)$  and $U_L(t)$, rescaled such that it tends to $1$ for $f\rightarrow 0$.
}
\label{fig_2D_spectre}

\end{figure}

\begin{figure}[htb!]
\begin{center}
\hbox{
\includegraphics[width=70mm,height=50mm]{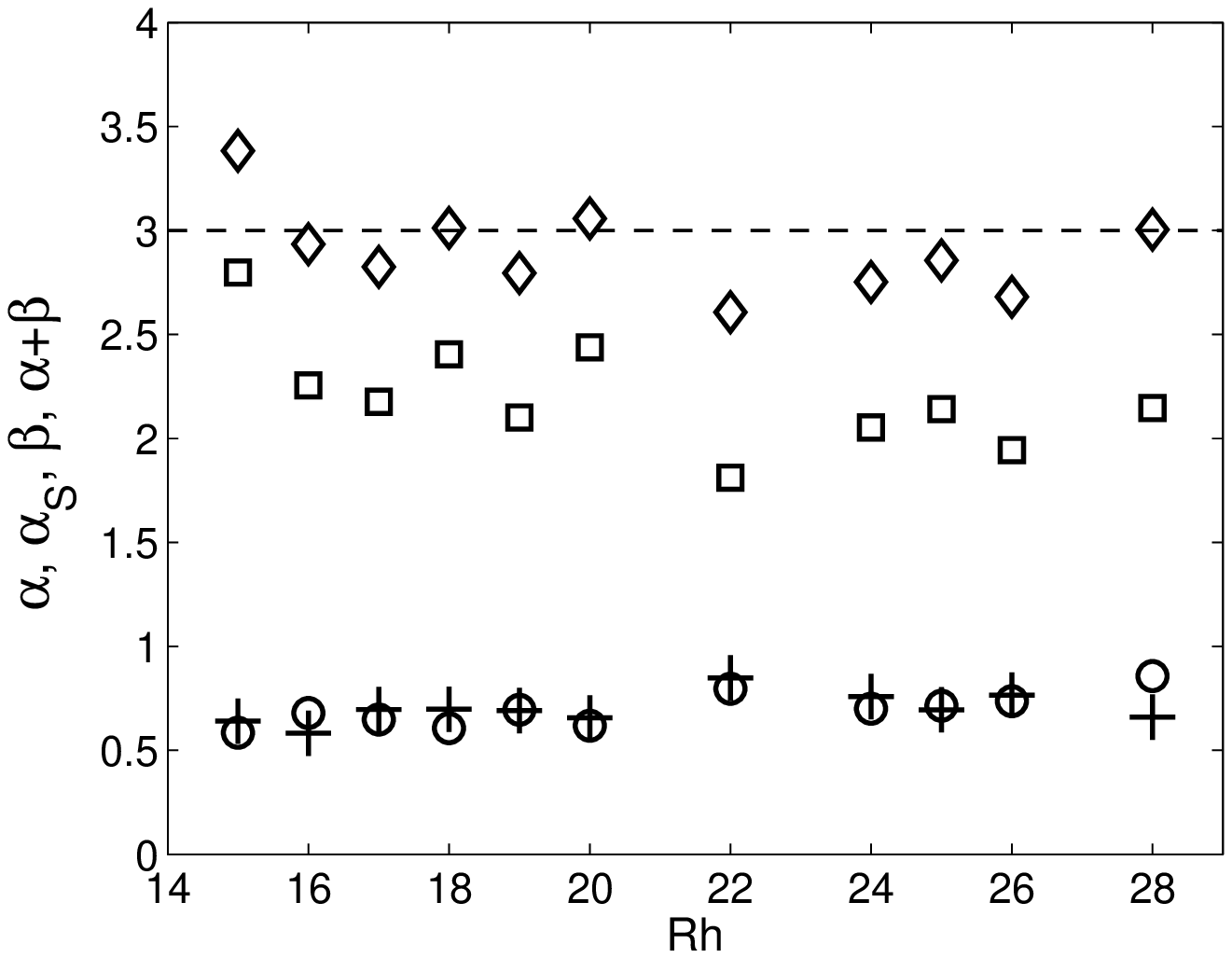}
\includegraphics[width=70mm,height=50mm]{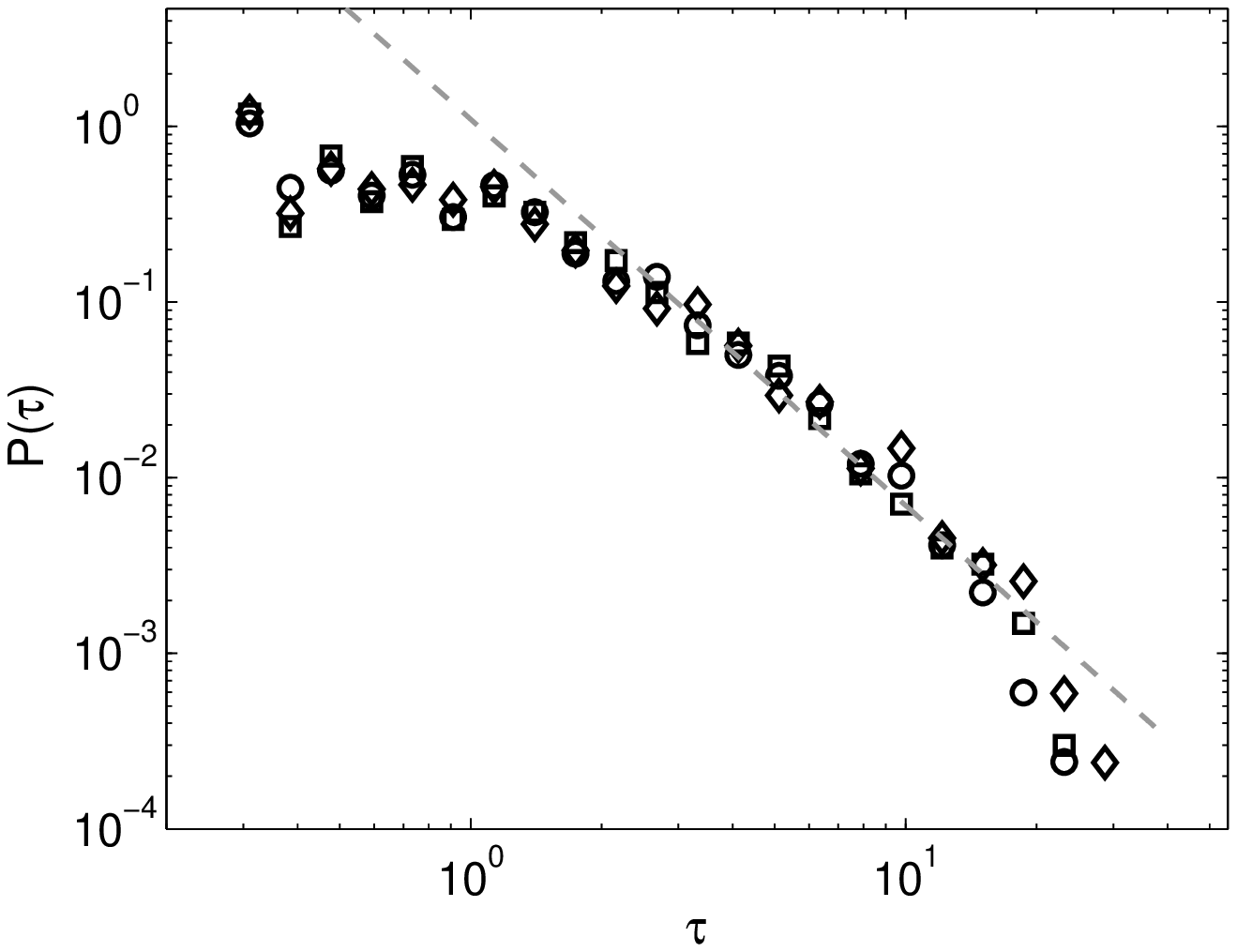}
}\caption{   
Left :  Exponents $\alpha$ (circle) and $\beta$ (square) for times series with different $Rh$. The exponent of the spectrum of $x(t)$ (plus)  and the sum $\alpha+\beta$ are  (diamond) also illustrated. Right : distribution $P(\tau)$ of the waiting time $\tau$ between sign change for $Rh=  16 ,   19  ,  24$. The dashed line  follows the law $\tau^{-2.25}$.
}

\label{fig_2D_exponents}
\end{center}
\end{figure}

These time series exhibit  $1/f$ power spectrum over one decade (Fig. \ref{fig_2D_spectre}) for $2.10^{-2} <f<2.10^{-1} Hz$. The   {related} frequencies are smaller than $\tau_D^{-1}$ the damping rate  of the  friction and smaller than the inverse  of the turn-over time $\tau_{L}=L/ \sigma_{U}$, with $\sigma_{U}$ the standard deviation of $U_L$. In  the context of 2D turbulence, $\tau_{L}^{-1}$  corresponds to the smallest frequency of the turbulent cascade. Thus the observed $1/f$ spectrum is not directly related to the energy cascade, but to the coherent dynamics of the large scale circulation. A systematic study  of the exponents $\alpha$, defined by $E(f) \sim f^{-\alpha}$,  shows that  {their} value is almost constant and equal to $0.7$ for   $Rh \in [15,28]$  (black circles, Fig. \ref{fig_2D_exponents}).

A closer inspection shows that the time series $U_L$ is composed of long phases of constant polarity of duration $\tau$ superimposed with random fluctuations. In  Fig. \ref{fig_2D_turb}, we highlight a phase of  constant polarity  with duration $\tau$ of $20$ s  {(see the event indicated by a double-arrow)}. The inverse of this  time $\tau^{-1}$ matches with the range of frequency of the $1/f$ spectrum.  Due to the symmetry of the forcing, $U_L$ displays random transitions between both polarities.  In order to show that these transitions generate  the $1/f$ noise, we decompose the signal $U_L$ into a two-state signal $s(t)$, which is defined by the sign of $U_L$, \textit{i.e.} $s\equiv\hbox{sign} \left(U_L \right)$. A typical time series $s$ is composed of one thousand transitions from one direction of rotation to the other. The ratio of the power spectrum of $U_L$ and of $s$ is reported in fig. \ref{fig_2D_spectre} and is almost constant over the frequency range corresponding to the $1/f$ spectrum. Thus both power spectra have the same behaviour at low frequency. A systematic study of the exponent $\alpha_s$ of the power spectrum of $s(t)$ (black cross in fig. \ref{fig_2D_exponents}) shows that its value is very close to the exponent $\alpha$ (black circles). This  implies that the slow dynamics of $U_L$, which is responsible for the $1/f$ noise, is mostly contained in $s(t)$.

We  calculate  the  probability density function  of the duration between sign changes of $s$ and report it in  Fig. \ref{fig_2D_exponents} (right). The distributions exhibit a power-law behavior on a range of durations $\tau \in [2,40] s$  corresponding to the inverse of the range of  frequencies $f \in [2.10^{-2},2.10^{-1}]$ Hz of the $1/f$ noise. We recover the association of a self-similar distribution of interevent duration with a self-similar spectrum of the time series. A systematic measurement of the exponent  $\beta$ (black square, fig. \ref{fig_2D_exponents}), defined by $P(\tau) \sim \tau^{-\beta}$, shows that it fluctuates around the value $\beta=2.25$. 

The sum $\alpha+\beta$ (diamonds in fig. \ref{fig_2D_exponents}) remains close to   $3$, as  predicted by equation (\ref{symAlpha}) for a symmetric renewal process.   All the previous results converge to the conclusion that the random changes of the sign of $U_L$ with waiting time distributed as  $P\sim \tau^{-2.25}$, is at the origin of the observed $1/f$ noise.

\subsection{Dynamics of the magnetic field generated by a von K\'arm\'an flow of liquid sodium}

 {The generation of magnetic field by a turbulent flow of liquid sodium has been widely studied in the VKS experiment, described in detail in ref. \cite{Monchaux}}. The flow is driven by two counter-rotating coaxial impellers in a cylindrical vessel (see fig. \ref{fig_VK_signal}). When the rotation rate of the impellers is larger than  a critical frequency $F_c$, a magnetic field  is generated by a self-sustained induction process, called the dynamo instability. The large scale magnetic field is mostly a stationary axial dipole  {with} superimposed magnetic fluctuations due to the strong turbulence of the flow.
 {$1/f$ noise in the power spectrum of the magnetic field has been reported in dynamo regimes \cite{Monchaux} and also below the dynamo threshold when an external magnetic field is imposed to the turbulent flow of liquid sodium \cite{Bourgoin}. We consider below the dynamo case but the same analysis hold for both.

\begin{figure}[htb!]
\begin{center}
\hbox{
\includegraphics[width=80mm,height=50mm]{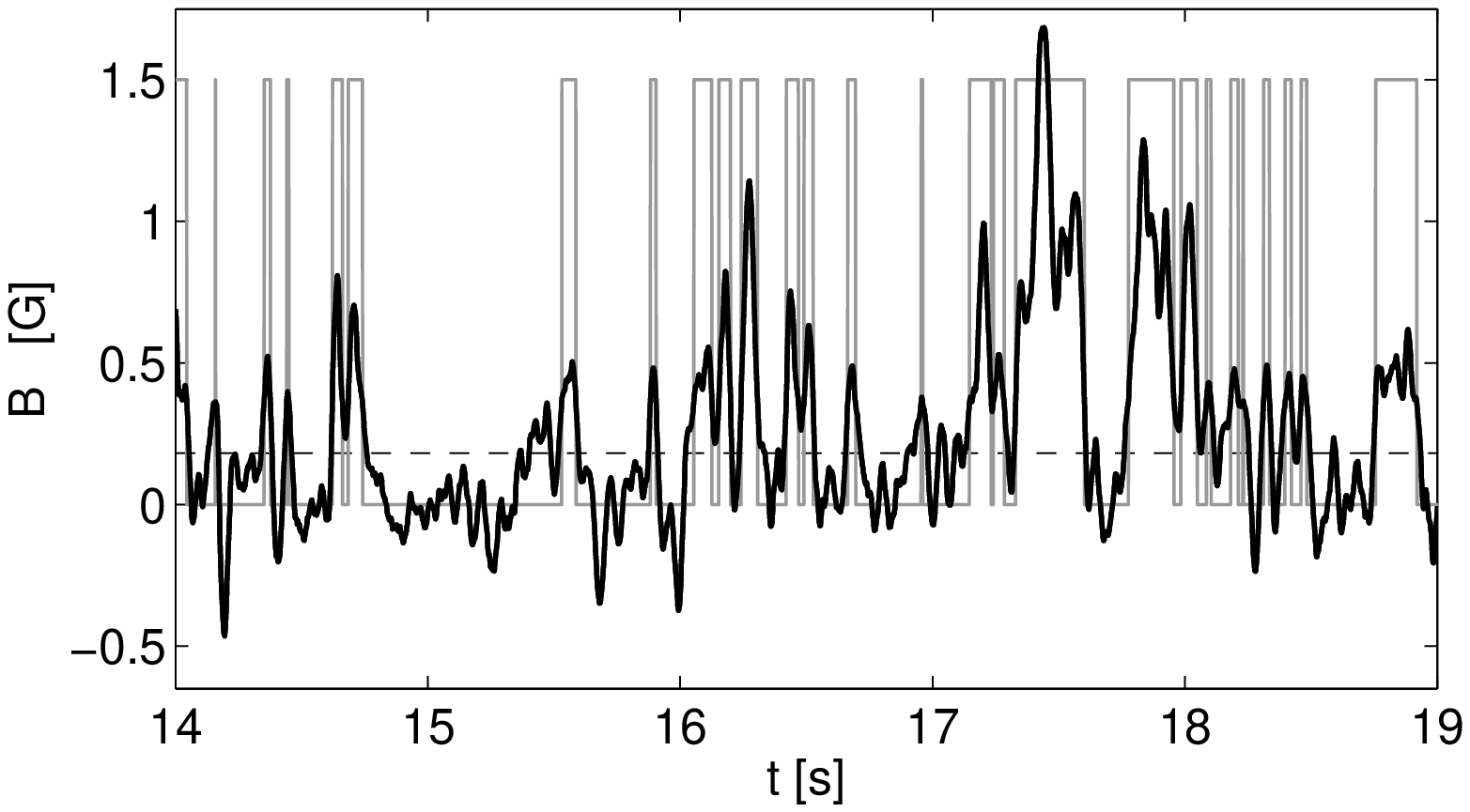}
\includegraphics[width=80mm,height=50mm]{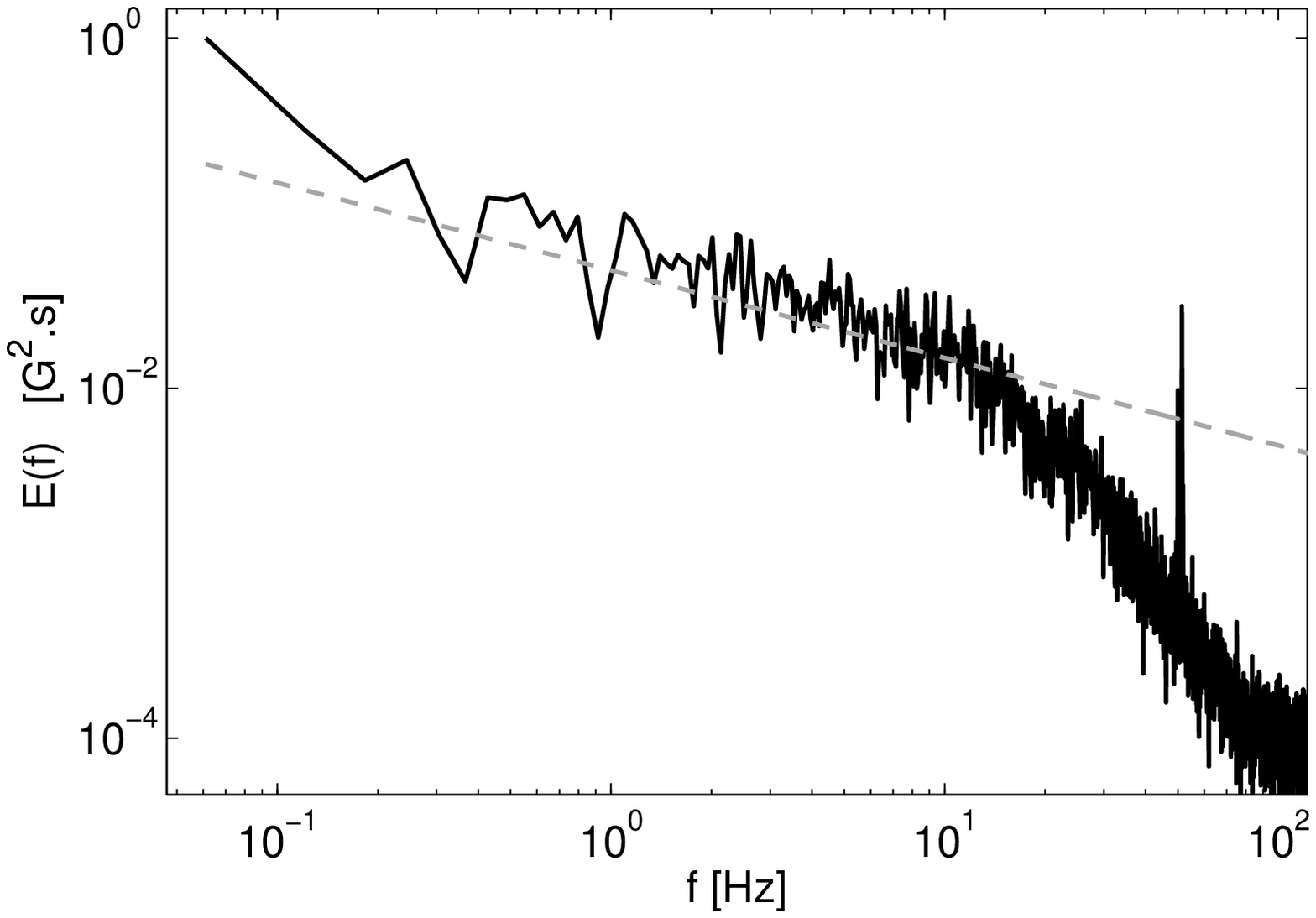}
}
\end{center}
\caption{Left: time series $B(t)$  of the azimuthal magnetic field, fluctuating around its mean value (dashed line). The grey line corresponds to the bursting process $s(t)$ extracted from $B(t)$. Right: power spectrum of $B(t)$ displaying a $f^{-0.5}$ behavior at low frequency (dashed line).}
\label{fig_VKS_spectre}

\end{figure}  

Measurements of the azimuthal magnetic field $B(t)$ are performed inside the vessel in the mid-plane between the impellers for $F=20$ Hz ( {just above the dynamo threshold}).  {The measurements} of other components of  magnetic field show similar results. A time series of $B(t)$ is displayed  in fig. \ref{fig_VKS_spectre} together with  the power spectrum which exhibits a clear $f^{-\alpha}$ spectrum with $\alpha \simeq 0.5$ for $f \in [1.5,15] Hz $.  For $f>20$ Hz, the power-spectrum scales as $f^{-11/3}$, due to the passive streching of the magnetic field by  the small scale turbulent fluctuations.

\begin{figure}[htb!]
\begin{center}
\hbox{
\includegraphics[width=80mm,height=50mm]{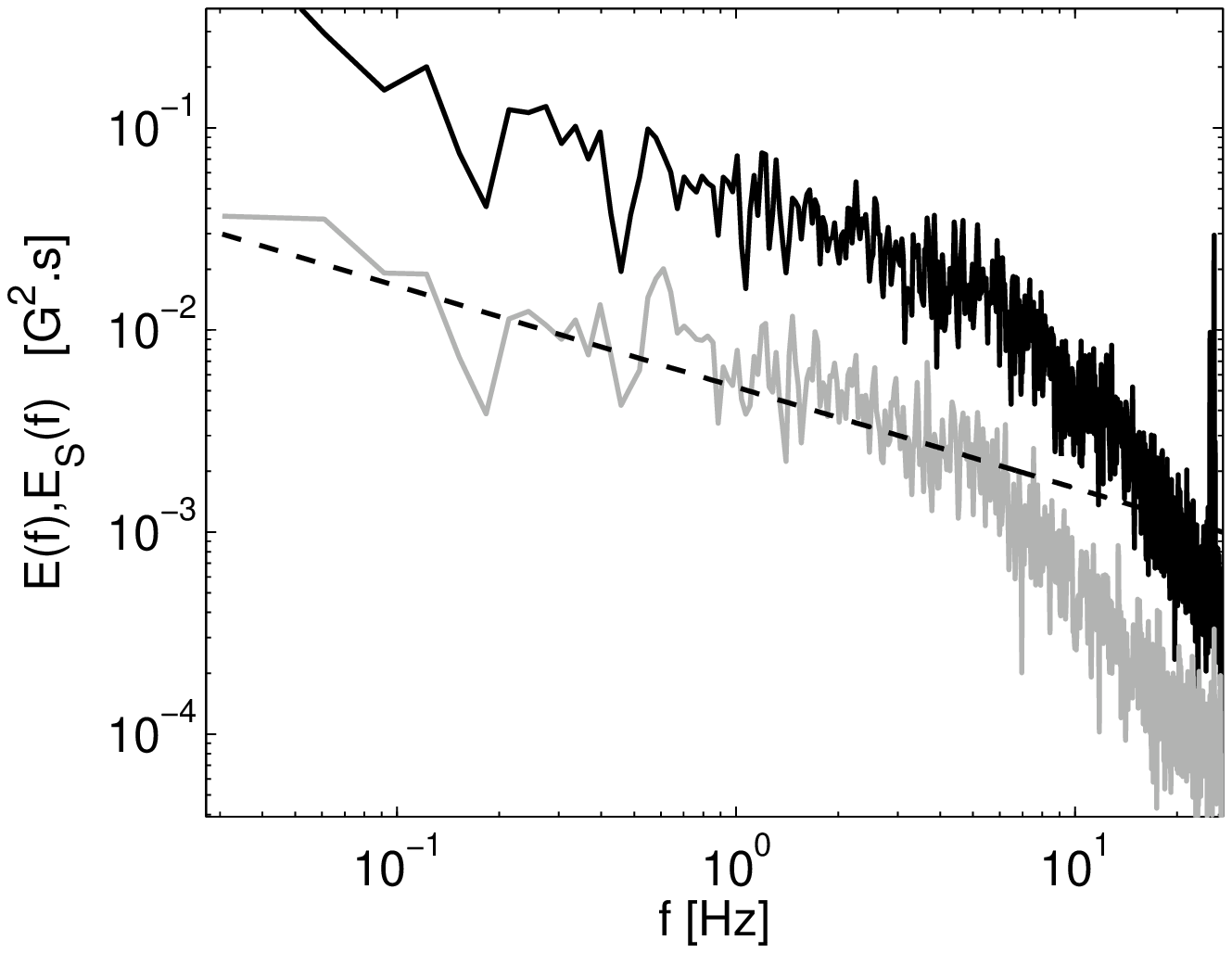}
\includegraphics[width=80mm,height=50mm]{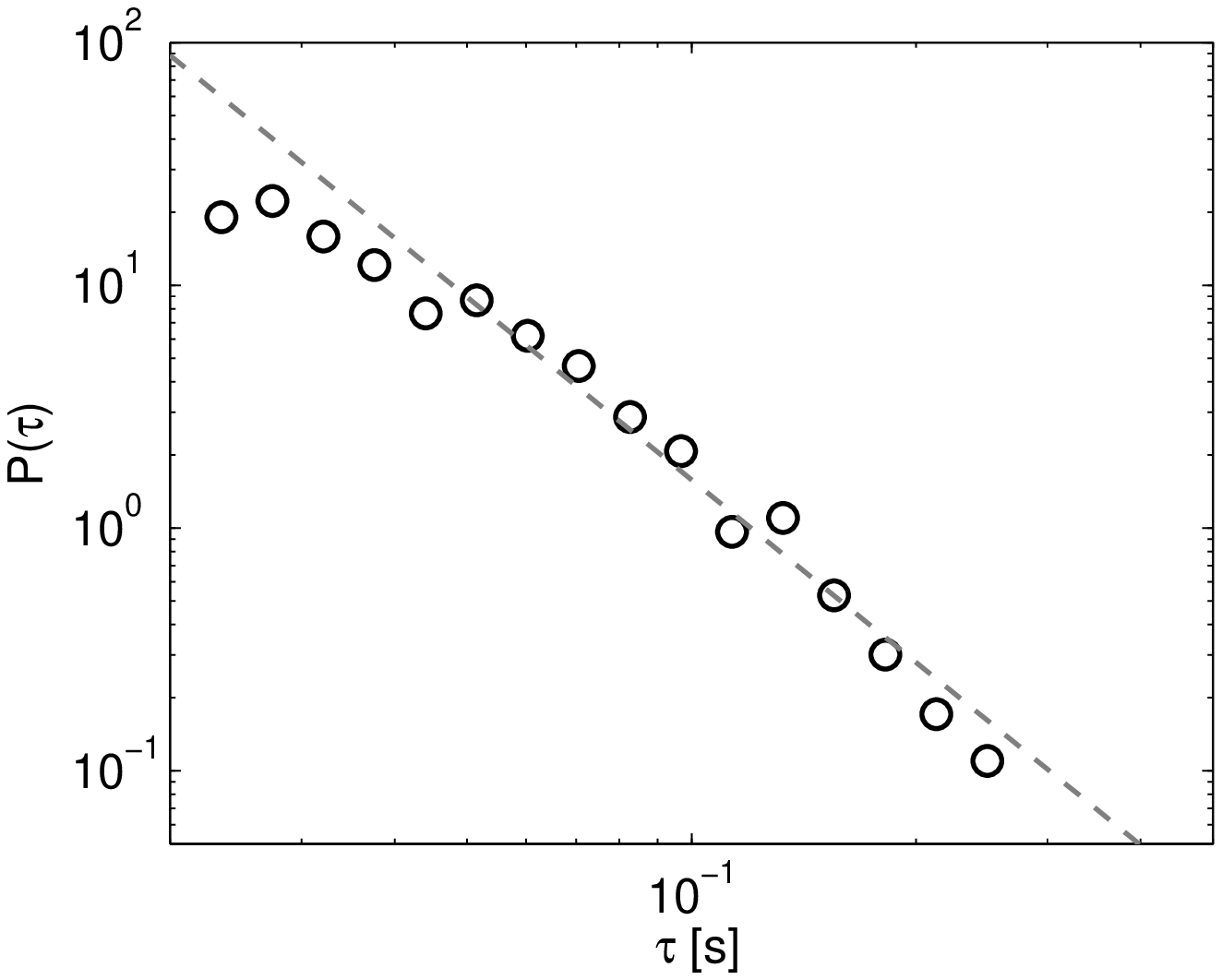} }
\caption{Left : power spectrum of $B(t)$ (black line) and $s(t)$ (grey line) compared to the law $f^{-0.6}$ (dashed line). Right : distribution of the waiting time $ \tau$  between bursts, with the law $\tau^{-2.5}$ (dashed line).}
\label{fig_VKS_spectre_s}
\end{center}
\end{figure}  

As observed in  fig. \ref{fig_VKS_spectre}, the magnetic field $B(t)$ displays bursts  {with}  amplitudes  {up to} $8$ times the average value. We thus define a  two-states signal $s(t)$  given by  phases of weak and large amplitude. Among the possible criteria to define a burst, we  {consider a} threshold  { larger than} the average value of $B$, such that above (resp. below) it, the system is in the high (resp. low) amplitude  {state}. The resulting  two-states signal $s(t)$ is displayed in fig \ref{fig_VKS_spectre} (left) in grey.  The power-spectrum of $s(t)$ (in grey) is compared to the one of $B$ (in black) in fig. \ref{fig_VKS_spectre_s} (left). At low frequency, both power spectra follow the same power-law. The distribution $P(\tau)$ of waiting times $\tau$ between bursts is displayed in fig. \ref{fig_VKS_spectre_s} (right). For the range of duration $\tau \in [5.10^{-2},25.10^{-2}]s$, $P(\tau)$ follows a power-law $\tau^{-\beta}$ with $\beta\simeq2.5$. 

We know from  equation (\ref{asymAlpha}) holding for burst processes, that $\alpha$ should be equal to $3-\beta$. The bursting process observed in the VKS experiment follows this relation with $\alpha=0.5$ and $\beta=2.5$.


\subsection{  Pressure fluctuations in von K\'arm\'an swirling flows }

 {$1/f$ noise has been also reported for the pressure signal in von K\'arm\'an swirling turbulent flows in water \cite{Abry}}. The results presented below are extracted from this reference where a detailed description of the experiment can be found. The pressure $p(t)$ is measured on the lateral  {boundary} of the cylinder.  Pressure drops are observed that are due to vorticity concentrations passing close to the pressure probe.

\begin{figure}[htb!]
\begin{center}
\hbox{
\includegraphics[width=80mm,height=50mm]{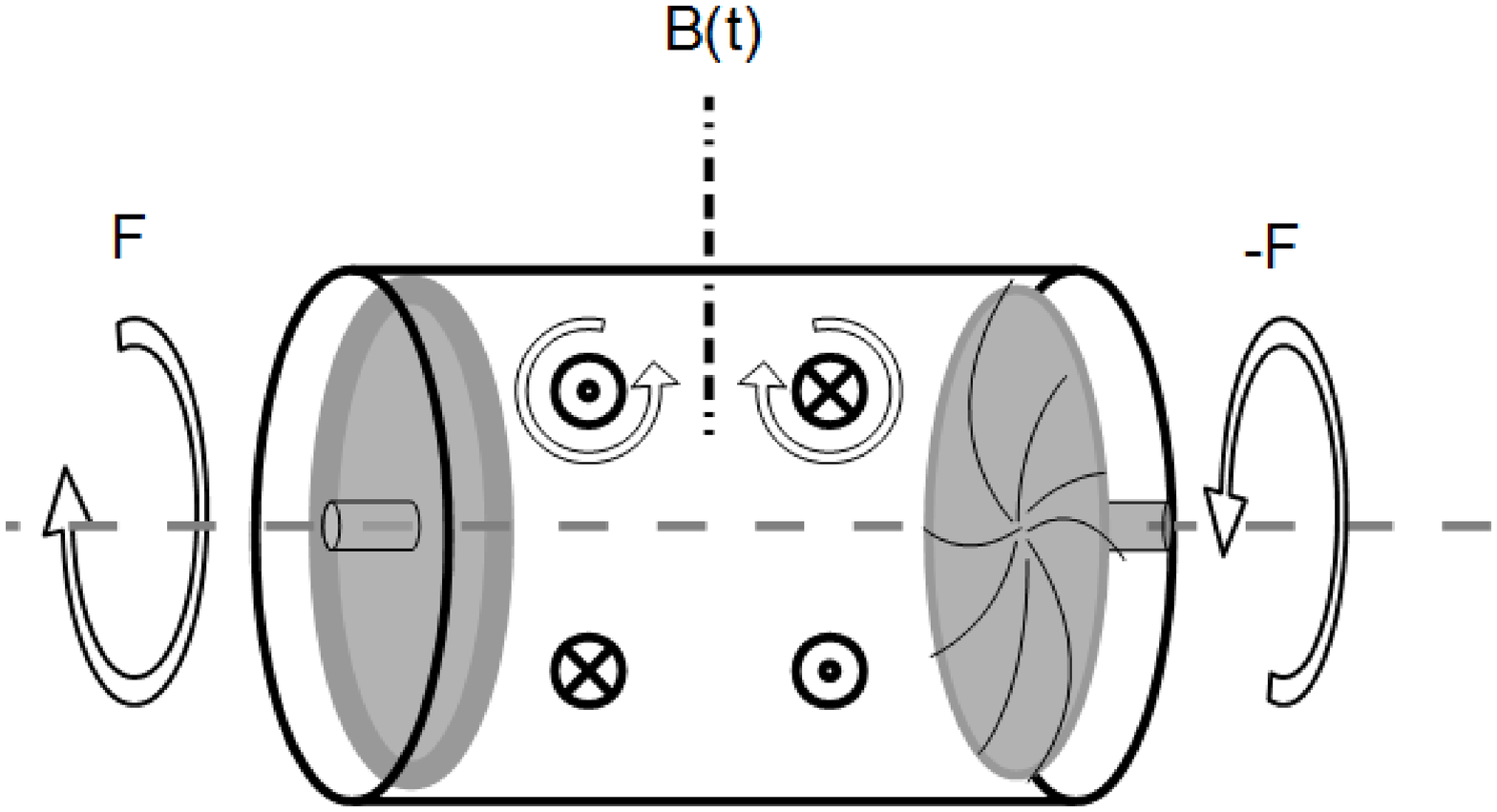}
\includegraphics[width=70mm,height=50mm]{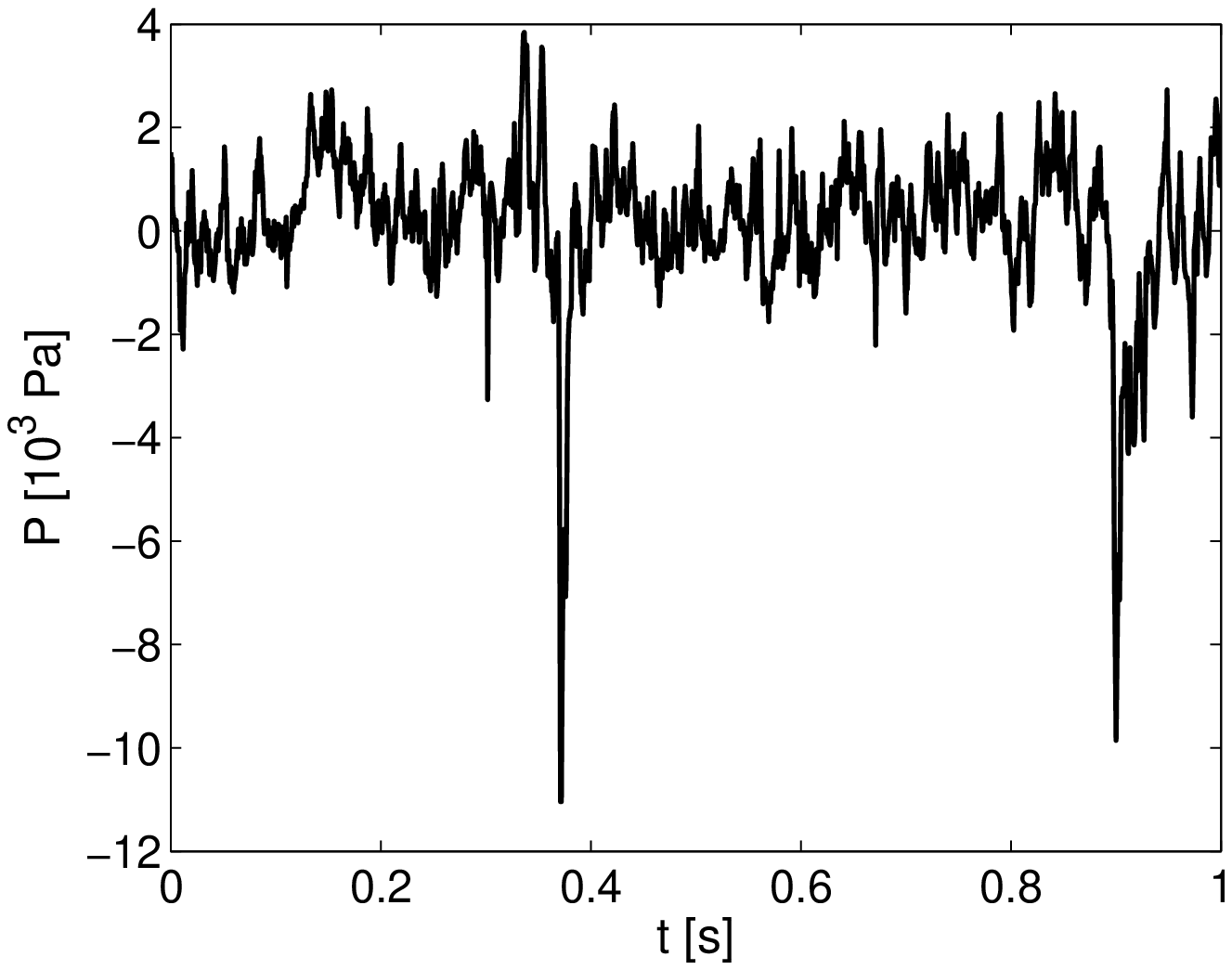}
}
\end{center}
\caption{ Left : Sketch of the VKS  and von K\'arm\'an experiments : two impellers counter-rotate at frequency $F$. The vertical thick line indicates the location of the magnetic probes in the VKS experiment. Right : time series $p(t)$ of the wall pressure from Abry et al. \cite{Abry}.  }
\label{fig_VK_signal}

\end{figure} 

The time series of $p(t)$ displaying large pressure drops is shown in fig. \ref{fig_VK_signal} (right). The  {related} power-spectrum is shown in fig. \ref{fig_VK_pdf} (left, black circles) and exhibits $1/f$ noise at low-frequency with an exponent $\alpha=0.6$.  {It has been shown in ref. \cite{Abry} that removing the pressure drops from the signal using a wavelet technique, almost suppresses the $1/f$ noise from the power-spectrum of the filtered signal $\tilde p$ 
(fig. \ref{fig_VK_signal}, left, black diamonds).  We observe again in this example that the $1/f$ noise results from the bursts. Thus, we define a  two-state signal   {that consists of} the long phases of weak amplitude between successive bursts and the short phases of large amplitude  {pressure drops} (the bursts). The distribution of  {waiting times} between successive bursts is displayed in fig. \ref{fig_VK_pdf} (right) and exhibits a power-law  {in} the range $\log(\tau) \in [-1.7,-0.5]$ ({\it i.e.} $\tau \in [0.02,0.3]$ s) corresponding to the frequency range $\log(f) \in[0.6,1.8]$ ({\it i.e.} $f \in [4,60]$ Hz) of the $1/f$ noise.  {In} this range, the fitted exponent is $\beta=1.58$. We also remark that the distribution has an exponential tail, but only for durations larger than the inverse of the minimum frequency over which $1/f$ noise is observed.

In contrast to  the previous case, the exponent $\beta$ is smaller than $2$ and the theoretical prediction is then $\alpha=\beta-1$. Once again, the agreement between the prediction and the fitted exponents is very good with $\beta-1 =0.58$ and $\alpha=0.6$. 

\begin{figure}[htb!]
\begin{center}
\hbox{
\includegraphics[width=70mm,height=50mm]{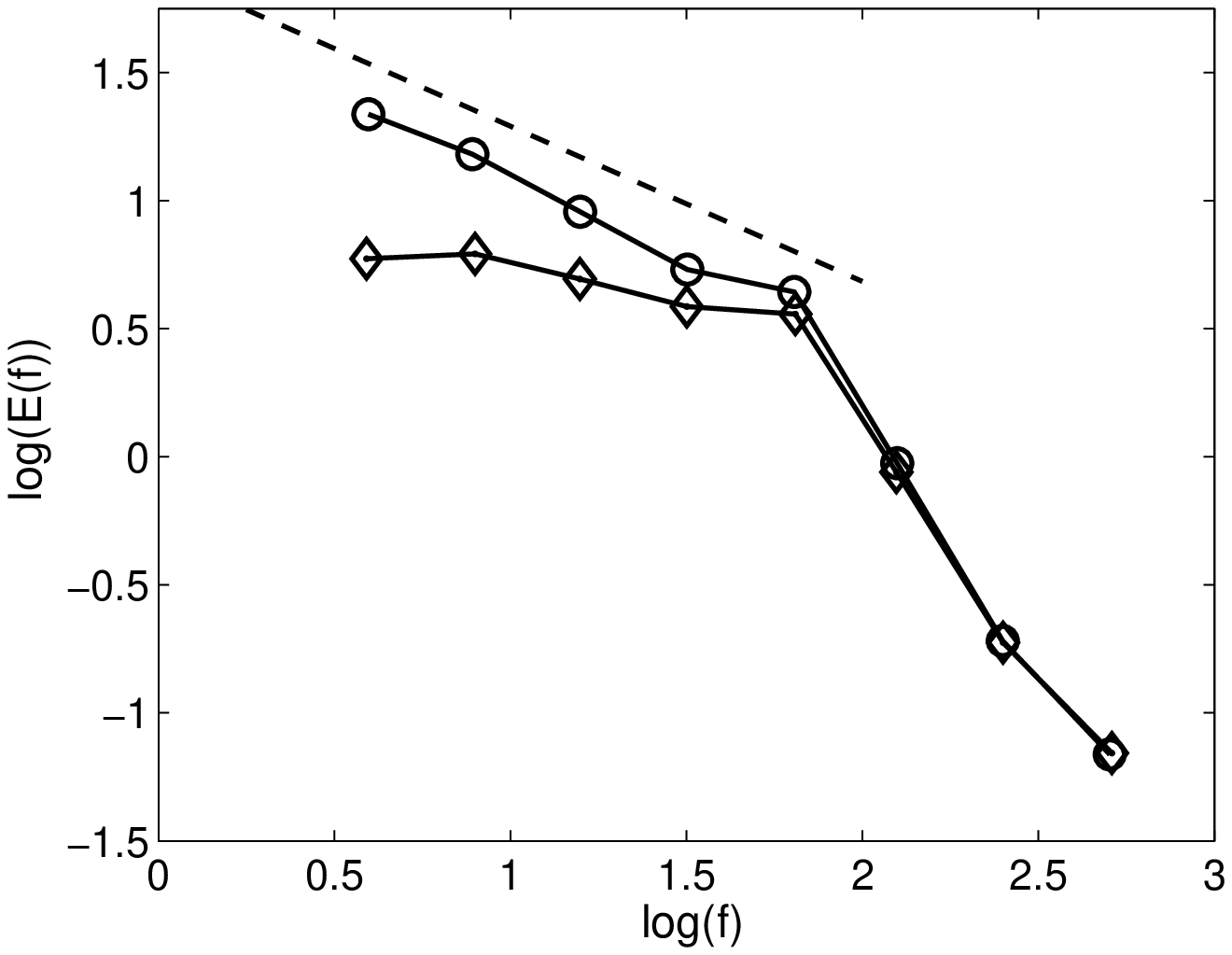}
\includegraphics[width=70mm,height=50mm]{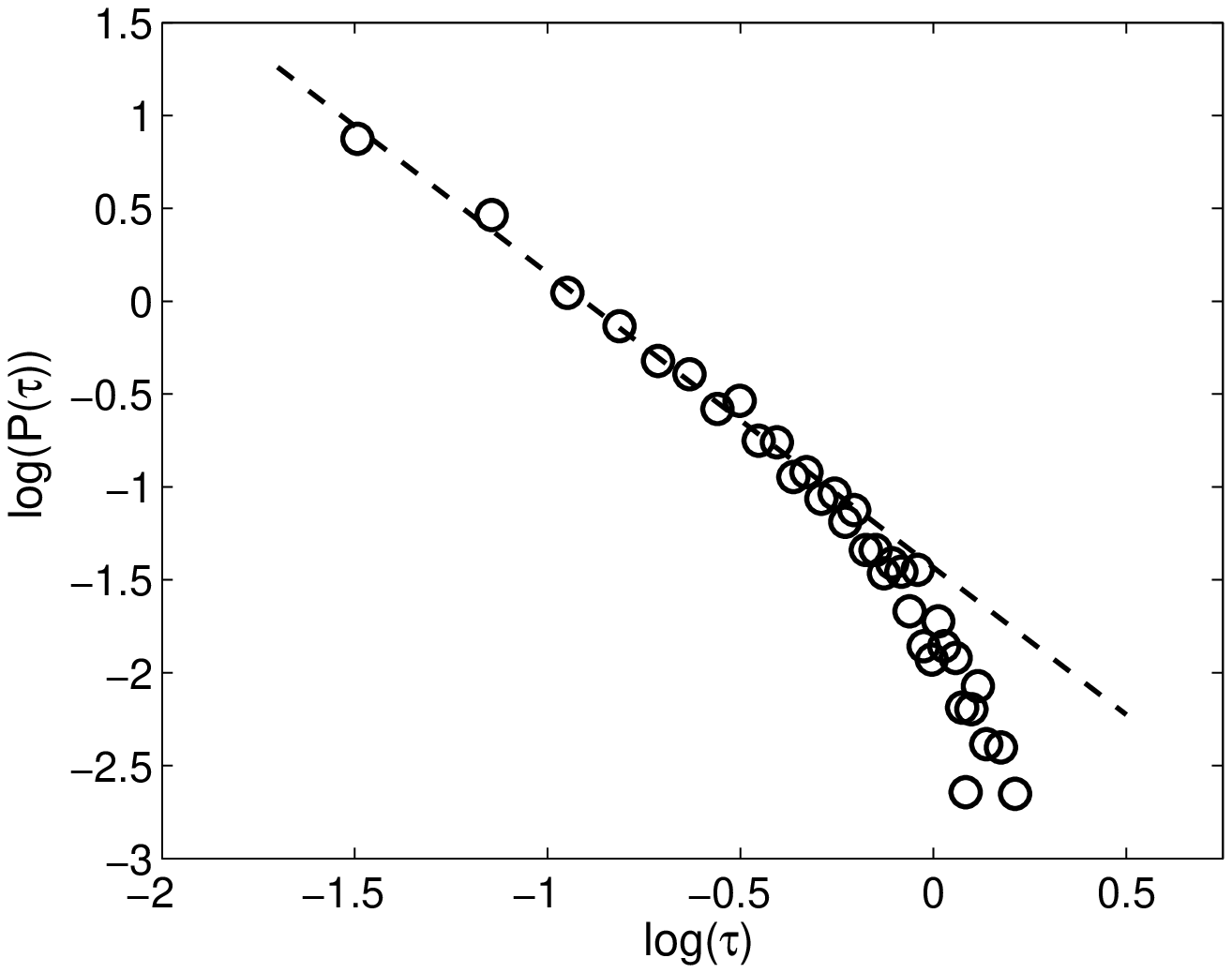}
}
\caption{Left : power spectra of $p(t)$ (circle) and $\tilde p$ (diamonds), the filtered pressure without bursts. The dashed line indicates the law $f^{-0.6}$. Right : distribution of the waiting time between pressure drops. The law $\tau^{-1.58}$ is displayed with a dashed line.}
\label{fig_VK_pdf}
\end{center}
\end{figure}

\section{Conclusion}

We have shown that the $1/f$ fluctuations observed experimentally in three different turbulent flows are  {related} to the coherent dynamics of large scale structures, which randomly transition between  {different} states. The dynamics is characterized by the nature of the process (asymmetric  bursts or symmetric transitions), and by the power-law distribution of the interevent durations. These two properties fully determine the exponent of the power spectrum.

\begin{acknowledgements}
The  authors
acknowledge their colleagues of the VKS team with whom the
experimental data have been obtained.
\end{acknowledgements}



\end{document}